\documentclass[twocolumn,aps,prc,showpacs,10pt]{revtex4-1}

\usepackage{amsmath}
\usepackage{amssymb}
\usepackage{graphicx}
\usepackage{mathrsfs}

\renewcommand{\vec}[1]{\mbox{\boldmath $#1$}}

\begin{document}
\title{Impurity effect of $\Lambda$ particle on the structure of 
$^{18}$F and $^{19}_{~\Lambda}$F}

\author{Y. Tanimura}
\author{K. Hagino}
\affiliation{Department of Physics, Tohoku University, Sendai 980-8578, Japan}

\author{H. Sagawa}
\affiliation{
Center for Mathematics and Physics, University of Aizu,
Aizu-Wakamatsu, Fukushima 965-8560, Japan}
\affiliation{
RIKEN Nishina Center, Wako 351-0198, Japan}

\date{\today}

\begin{abstract}
We perform three-body model calculations for 
a $sd$-shell hypernucleus $^{19}_{\ \Lambda}$F 
($^{17}_{\ \Lambda}{\rm O}+p+n$) and its core nucleus $^{18}$F ($^{16}{\rm O}+p+n$), 
employing 
a density-dependent contact interaction between the valence proton and neutron. 
We find that the $B(E2)$ value from the first excited 
state (with spin and parity of $I^\pi=3^+$) to the ground state ($I^\pi=1^+$) 
is slightly decreased by the addition of a $\Lambda$ particle, 
which exhibits the so called shrinkage effect of 
$\Lambda$ particle. 
We also show that the excitation energy of the $3^+$ state is reduced in 
$^{19}_{\ \Lambda}$F compared to $^{18}$F, as is observed in a $p$-shell nucleus $^{6}$Li. 
We discuss the mechanism of this reduction of the excitation energy, 
pointing out that it is caused by a different mechanism 
from that in $^{7}_{\Lambda}$Li. 
\end{abstract}


\maketitle

\begin{section}{INTRODUCTION}
It has been of great interest in hypernuclear physics to investigate 
how $\Lambda$ particle affects the core nucleus when it is added to 
a normal nucleus.  
A $\Lambda$ particle may change various 
nuclear properties, e.g., nuclear size and shape \cite{MBI83,MHK11,LZZ11,IsKDO11p,IsKDO12}, 
cluster structure \cite{IsKDO11sc}, the neutron drip line \cite{VPLR98,ZPSV08}, the fission barrier \cite{MiCH09}, and the collective 
excitations \cite{Yao11,MH12}. 
Such effects caused by $\Lambda$ 
on nuclear properties are 
referred to as an 
impurity effect. Because $\Lambda$ particle can penetrate deeply into a nucleus without 
the Pauli principle from nucleons, a response of the core nucleus to an addition of a $\Lambda$ 
may be essentially different from that to non-strange probes. That is, $\Lambda$ particle can 
be a unique probe of nuclear structure that cannot be investigated 
by normal reactions. 

The low-lying spectra and electromagnetic transitions have been measured 
systematically in $p$-shell hypernuclei by high precision $\gamma$-ray spectroscopy \cite{HT06}. 
The experimental data have indicated 
a shrinkage of nuclei due to the attraction of $\Lambda$. 
A well-known example is $^7_{\Lambda}$Li, for which the electric quadrupole transition 
probability, $B(E2)$, from the first excited state ($3^+$) to the ground state ($1^+$)
of $^6{\rm Li}$ is considerably reduced when a $\Lambda$ particle is added \cite{Ta01,HiKMM99}. 
This reduction of the $B(E2)$ value has been interpreted as a shrinkage of the distance 
between $\alpha$ and $d$ clusters in $^{6}$Li. 
On the other hand, a change of excitation energy 
induced by a $\Lambda$ particle 
depends on nuclides. 
If one naively regards a di-cluster nucleus as a classical rigid rotor, 
shrinkage of nuclear size would lead to 
a reduction of the moment of inertia, increasing 
the rotational excitation energy. 
However, $^6$Li and $^8$Be show 
a different behavior from this naive expectation. 
That is, the spin averaged excitation 
energy decreases in $^6$Li \cite{Uk06} while it 
is almost unchanged in $^8$Be \cite{Ak02}. 

Recently Hagino and Koike \cite{HK11} have applied a semi-microscopic cluster model to $^6$Li, 
$^{7}_{\Lambda}$Li, $^8$Be, and $^9_{\Lambda}$Be to successfully account for the relation between 
the shrinkage effect and the rotational spectra 
of the two nuclei simultaneously. 
They argue that a Gaussian-like potential  between two clusters leads to a stability of excitation 
spectrum against an addition of a $\Lambda$ particle, 
even though the intercluster distance is reduced. 
This explains the stabilization of the spectrum in $^8$Be. In the case of lithium one has to consider also 
the spin-orbit interaction between $^4$He/$^5_{\Lambda}$He and the deuteron cluster. 
Because of the shrinkage effect of $\Lambda$, the overlap between 
the relative wave function and the spin-orbit potential 
becomes larger in $^7_{\Lambda}$Li than in $^6$Li. 
This effect lowers the 
$3^+\otimes\Lambda_{s_{1/2}}$ state 
more than the 3$^+$ state in $^6$Li, making 
the rotational excitation energy in $^7_{\Lambda}$Li smaller than in $^6$Li. 

These behaviors of the spectra may be specific to 
the two-body cluster structure. 
$^6$Li and $^8$Be have in their ground states well-developed $\alpha$ cluster structure. 
In heavier nuclei, on the other hand, 
cluster structure appears in their excited states while the ground and low-lying states 
have a mean-field-like structure. 
In this respect, 
it is interesting to investigate 
the impurity effect on 
a $sd$-shell nucleus $^{18}$F, 
in which the mean-field structure and $^{16}{\rm O}+d$ 
cluster structure may be mixed\cite{SNN76,LMTT76,ISHA67,BMR79,BFP77,M83}. 
Notice that 
the ground and the first excited states of $^{18}$F are 
$1^+$ and $3^+$, respectively, which are the same as $^6$Li. 
We mention that 
a $\gamma$-ray spectroscopy measurement for $^{19}_{~\Lambda}$F 
is planned at J-PARC facility 
as the first $\gamma$-ray experiment for $sd$-shell 
hypernuclei \cite{JPARC,Sendai08}. 

In this paper we 
employ a three-body model of $^{16}{\rm O}+p+n$ for $^{18}$F 
and of $^{17}_{\ \Lambda}{\rm O}+p+n$ for $^{19}_{\ \Lambda}$F, and 
study the structure change of $^{18}$F caused by the impurity 
effect of a $\Lambda$ particle. 
This model enables us to 
describe both 
mean-field and cluster like structures of these nuclei.  
We discuss how $\Lambda$ particle affects the electric transition 
probability $B(E2,3^+\to1^+)$, the density distribution of the valence nucleons, 
and the excitation energy. Of particular interest is whether 
the excitation energy 
increases or decreases due to the $\Lambda$ particle. 
We discuss the mechanism of its change in comparison with the lithium nuclei. 

The paper is organized as follows. In Sec. \ref{sec:model}, 
we introduce the three-body model 
to describe $^{18}$F and $^{19}_{\ \Lambda}$F. 
In Sec. \ref{sec:result}, we present the results and discuss the relation between the shrinkage effect 
and the energy spectrum. 
In Sec. \ref{sec:summary}, we summarize the paper. 
\label{sec:intro}
\end{section}

\begin{section}{THE MODEL}

\begin{subsection}{Hamiltonian}
We employ a three-body model to describe 
the $^{18}$F and $^{19}_{~\Lambda}$F nuclei. 
We first describe the model Hamiltonian for the $^{18}$F nucleus, assuming 
the $^{16}$O + $p + n$ structure. 
After removing the center-of-mass 
motion, it is given by 
\begin{eqnarray}
H&=&\frac{\vec{p}_p^2}{2m}+\frac{\vec{p}_n^2}{2m}+V_{pC}(\vec{r}_p)+V_{nC}(\vec{r}_n) \nonumber \\
&&+V_{pn}(\vec{r}_p,\vec{r}_n)+\frac{(\vec{p}_p+\vec{p}_n)^2}{2A_{C}m},
\label{eq:H}
\end{eqnarray}
where $m$ is the nucleon mass and $A_C$ is the mass number of the core nucleus. 
$V_{pC}$ and $V_{nC}$ is the mean field potentials 
for proton and 
neutron, respectively, 
generated by the core nucleus. 
These are given as 
\begin{eqnarray}
V_{nC}(\vec{r}_n)=V^{(N)}(r_n),\ V_{pC}(\vec{r}_p)=V^{(N)}(r_p)+V^{(C)}(r_p), 
\label{eq:pot}
\end{eqnarray}
where $V^{(N)}$ and $V^{(C)}$ are the nuclear and the Coulomb parts, respectively. 
In Eq. (\ref{eq:H}), $V_{pn}$ is the interaction between the two valence 
nucleons. 
For simplicity, we neglect in this paper 
the last term in Eq. (\ref{eq:H}) since the core $^{16}$O is much 
heavier than nucleons. 
Then the Hamiltonian reads
\begin{eqnarray}
H=h(p)+h(n)+V_{pn},
\label{eq:Htot}
\end{eqnarray}
where the single-particle Hamiltonians are given as  
\begin{eqnarray}
h(p)=\frac{\vec{p}_p^2}{2m}+V_{pC}(\vec{r}_p),\ 
h(n)=\frac{\vec{p}_n^2}{2m}+V_{nC}(\vec{r}_n).
\label{eq:hsp}
\end{eqnarray}

In this paper, the nuclear part of the mean-field potential, $V^{(N)}$, 
is taken to be a spherical Woods-Saxon type
\begin{eqnarray}
V_n(r)=\frac{v_0}{1+e^{(r-R)/a}}
+({\boldsymbol \ell}\cdot\vec{s})\frac{1}{r}\frac{d}{dr}
\frac{v_{\ell s}}{1+e^{(r-R)/a}}, 
\label{eq:WS}
\end{eqnarray}
where the radius and the diffuseness parameters are set to be 
$R=1.27A_C^{1/3}$ fm and $a=0.67$ fm, respectively, 
and the strengths $v_0$ and $v_{\ell s}$ are determined 
to reproduce the neutron single-particle 
energies of $2s_{1/2}$ ($-3.27$ MeV) and $1d_{5/2}$ ($-4.14$ MeV) orbitals in $^{17}$O \cite{TWC93}. 
The resultant values are $v_0=-49.21$ MeV and $v_{\ell s}=21.6\ {\rm MeV}\cdot
{\rm fm}^2$. 
The Coulomb potential $V^{(C)}$ in the proton mean field potential 
is generated by a uniformly 
charged sphere of radius $R$ and charge $Z_Ce$, where $Z_C$ is the atomic number of the core nucleus. 
For the pairing interaction $V_{pn}$ we employ a density-dependent contact interaction, which 
is widely used in similar three-body calculations 
for nuclei far from the stability line\cite{BeEs91,EsBeH97,HSCS07,OiHS10}. 
Since we have to consider both the iso-triplet and iso-singlet 
channels in our case of proton and neutron, 
we consider the pairing interaction $V_{pn}$ given by 
\begin{eqnarray}
&&V_{pn}(\vec{r}_p,\vec{r}_n) \nonumber \\
&=&\hat{P}_sv_s\delta^{(3)}(\vec{r}_p-\vec{r}_n)
\biggl[1+x_s\biggl(\frac{1}{1+e^{(r-R)/a}}\biggr)^{\alpha_s}\biggr]\nonumber \\
&+&\hat{P}_tv_t\delta^{(3)}(\vec{r}_p-\vec{r}_n)
\biggl[1+x_t\biggl(\frac{1}{1+e^{(r-R)/a}}\biggr)^{\alpha_t}\biggr],
\label{eq:pairing}
\end{eqnarray}
where $\hat{P}_s$ and $\hat{P}_t$ are the projectors onto the 
spin-singlet and spin-triplet channels, respectively: 
\begin{eqnarray}
\hat{P}_s=\frac{1}{4}-\frac{1}{4}{\boldsymbol \sigma}_p\cdot{\boldsymbol \sigma}_n,\ 
\hat{P}_t=\frac{3}{4}+\frac{1}{4}{\boldsymbol \sigma}_p\cdot{\boldsymbol \sigma}_n. 
\end{eqnarray}
In each channel 
in Eq. (\ref{eq:pairing}),  
the first term 
corresponds to the interaction in vacuum while 
the 
second term takes into account the medium effect through the density 
dependence. Here, the core density is assumed to be 
a Fermi distribution of the same radius and diffuseness as in 
the mean field, Eq. (\ref{eq:WS}).  
The strength parameters, $v_s$ and $v_t$ are determined from the proton-neutron scattering length as \cite{EsBeH97}
\begin{eqnarray}
v_s&=&\frac{2\pi^2\hbar^2}{m}\frac{2a^{(s)}_{pn}}{\pi-2a^{(s)}_{pn}k_{\rm cut}},\\
v_t&=&\frac{2\pi^2\hbar^2}{m}\frac{2a^{(t)}_{pn}}{\pi-2a^{(t)}_{pn}k_{\rm cut}}, 
\label{eq:v0pair}
\end{eqnarray}
where $a^{(s)}_{pn}=-23.749$ fm and $a^{(t)}_{pn}=5.424$ fm \cite{KoNi75} are 
the empirical p-n scattering length of the spin-singlet and spin-triplet 
channels, respectively, and $k_{\rm cut}$ is the momentum cut-off introduced in treating the 
delta function. The density dependent terms have two parameters, $x$ and $\alpha$, for each channel, which are 
to be determined so as to reproduce the ground and excited state 
energies of $^{18}$F (see Sec. III). 

\end{subsection}

\begin{subsection}{Model space}
The Hamiltonian, Eq. (\ref{eq:Htot}), is diagonalized in the valence two-particle subspace. 
The basis is given by a product of proton and neutron single-particle states:
\begin{eqnarray}
\begin{aligned}
h(\tau)\bigl|\psi^{(\tau)}_{n\ell jm}\bigr\rangle
=\epsilon_{n\ell j}^{(\tau)}\bigl|\psi^{(\tau)}_{n\ell jm}\bigr\rangle,\ 
\tau=p\ {\rm or}\ n,
\end{aligned}
\end{eqnarray}
where the single-particle continuum states can be discretized in a large box. 
Here, $n$ is the principal quantum number, $\ell$ is the orbital angular 
momentum, $j$ is the total angular momentum of the single-particle state, 
and $m=j_z$ is the projection of the total angular momentum $j$. 
$\epsilon_{n\ell j}^{(\tau)}$ is the single-particle energy. 
The wave function for states of the 
total angular momentum $I$ is expanded as 
\begin{eqnarray}
|\Psi_{IM_I}\rangle=\sum_{\alpha\beta}C_{\alpha\beta}^{I}|\alpha\beta,IM_I\rangle, 
\label{eq:wf3body}
\end{eqnarray}
where $C_{\alpha\beta}^{I}$ are the expansion coefficients. 
The basis state $|\alpha\beta,IM_I\rangle$ is given by the product 
\begin{eqnarray}
&&\langle\vec{r}_p\vec{r}_n|\alpha\beta,IM_I\rangle \nonumber \\
&=&\phi^{(p)}_{\alpha}(r_p)\phi^{(n)}_{\beta}(r_n)
[{\mathscr Y}_{\ell_{\alpha}j_{\alpha}}(\hat{\vec{r}}_p)
{\mathscr Y}_{\ell_{\beta}j_{\beta}}(\hat{\vec{r}}_n)]_{IM_I},
\label{eq:basis}
\end{eqnarray}
where $\alpha$ is a shorthanded notation for single-particle level 
$\{n_{\alpha},\ell_{\alpha},j_{\alpha}\}$, and similarly for $\beta$. 
$\phi^{(\tau)}_{\alpha}(r_{\tau})$ is the radial part of the wave function 
$\psi_\alpha^{(\tau)}$ of level 
$\alpha$, and 
${\mathscr Y}_{\ell jm}=
\sum_{m'm''}\langle \ell m' \frac{1}{2}m''|jm \rangle Y_{\ell m'}\chi_{\frac{1}{2}m''}$ 
is the spherical spinor, $\chi_{\frac{1}{2}m''}$ being the spin wave function of 
nucleon. 
$\ell_{\alpha}+\ell_{\beta}$ is even (odd) for positive (negative) parity state. 
Notice that we do not use the isospin formalism, with which 
the number of the basis states, Eq. (\ref{eq:basis}), can be reduced 
by explicitly taking the anti-symmetrization. Instead, we use the 
proton-neutron formalism without the anti-symmetrization in order to 
take into account the breaking of the isospin symmetry 
due to the Coulomb term $V^{(C)}$ in Eq. (\ref{eq:pot}). 

As shown in the Appendix A, 
the matrix elements of the spin-singlet channel in $V_{pn}$ 
identically vanish for the $1^+$ and $3^+$ states. 
Thus we keep only the spin-triplet channel interaction and determine 
$x_t$ and $\alpha_t$ from the binding 
energies of the two states from the three-body threshold. 
In constructing the basis we 
effectively take into account the Pauli principle, and 
exclude the single-particle $1s_{1/2}$, $1p_{3/2}$, and $1p_{1/2}$ 
states, that 
are already occupied by the core nucleons. 
The cut-off energy $E_{\rm cut}$ to truncate the model space are related with the momentum 
cut-off in Eq. (\ref{eq:v0pair}) by $E_{\rm cut}=\hbar^2k_{\rm cut}^2/m$. We include only 
those states satisfying $\epsilon^{(p)}_{\alpha}+\epsilon^{(n)}_{\beta}\leq E_{\rm cut}$.

\end{subsection}

\begin{subsection}{Addition of a $\Lambda$ particle}

Similarly to the $^{18}$F nucleus, 
we also treat $^{19}_{\ \Lambda}$F as a three-body system composed of 
$^{17}_{\ \Lambda}{\rm O}+p+n$. 
We assume that the $\Lambda$ particle 
occupies the $1s_{1/2}$ orbital in the core nucleus 
and provides an additional contribution to the core-nucleon potential, 
\begin{eqnarray}
V^{(N)}(r)\rightarrow V^{(N)}(r)+V_{\Lambda}(r).
\end{eqnarray} 
We construct the potential 
$V_{\Lambda}$ by folding 
the $N$-$\Lambda$ interaction $v_{N\Lambda}$ with 
the density $\rho_{\Lambda}$ of the $\Lambda$ particle: 
\begin{eqnarray}
V_{\Lambda}(r)=\int d^3r_{\Lambda}\ \rho_{\Lambda}(\vec{r}_{\Lambda})
v_{N\Lambda}(\vec{r}-\vec{r}_{\Lambda}). 
\end{eqnarray}
We use the central part of a $N$-$\Lambda$ 
interaction by Motoba {\it et al.} \cite{MBI83}: 
\begin{eqnarray}
v_{\Lambda N}(r)=v_{\Lambda}e^{-r^2/b_v^2}, 
\label{eq:vNL}
\end{eqnarray}
where $b_v=1.083$ fm and we set $v_{\Lambda}=-20.9$ MeV, 
which is used in the calculation for $^6$Li in Ref. \cite{HK11}. 
The density $\rho_{\Lambda}(r)$ is given by that of a harmonic oscillator 
wave function 
\begin{eqnarray}
\rho_{\Lambda}(r)=(\pi b_{\Lambda}^2)^{-3/2}e^{-r^2/b_{\Lambda}^2}, 
\end{eqnarray}
where we take $b_{\Lambda}=\sqrt{(A_C/4)^{1/3}(A_Cm+m_{\Lambda})/A_Cm_{\Lambda}}\cdot 1.358$ fm, 
following Refs. \cite{MBI83} and \cite{HK11}. 

The total wave function for the $^{19}_{~\Lambda}$F nucleus is given by 
\begin{equation}
|\Psi^{\rm tot}_{JM}\rangle = \left[|\Phi_{I_c}\rangle|\Psi_I\rangle\right]^{(JM)},
\end{equation}
where $J$ is the total angular momentum 
of the $^{19}_{~\Lambda}$F nucleus, 
$|\Phi_{I_c}\rangle$ is the wave function for the core nucleus, 
$^{17}_{~\Lambda}$O, 
in the ground state with the spin-parity of $I_c^\pi=1/2^+$, and 
$|\Psi_I\rangle$ is the wave function for the valence nucleons 
with the angular momentum $I$ given by Eq. (\ref{eq:wf3body}). 
As we use the spin-independent $N$-$\Lambda$ interaction in 
Eq. (\ref{eq:vNL}), the doublet states with $J=I\pm 1/2$ are degenerate 
in energy.

\end{subsection}

\begin{subsection}{$E2$ transition and the polarization charge}
We calculate the reduced electric transition probability, $B(E2,3^+\to1^+)$, as a 
measure of nuclear size. In our three-body model, the 
$E2$ transition operator $Q_{2\mu}$ is given by
\begin{eqnarray}
&&Q_{2\mu} \nonumber \\
&=&\frac{(Z_Ce+e^{(n)})m^2+e^{(p)}(M_C+m)^2}{(M_C+2m)^2}r_p^2Y_{2\mu}(\hat{\vec{r}}_p) \nonumber \\
&+&\frac{(Z_Ce+e^{(p)})m^2+e^{(n)}(M_C+m)^2}{(M_C+2m)^2}r_n^2Y_{2\mu}(\hat{\vec{r}}_n) \nonumber \\
&+&2\frac{Z_Cem^2-(e^{(p)}+e^{(n)})m(M_C+m)}{(M_C+2m)^2}\nonumber \\
&&\times\sqrt{\frac{10\pi}{3}}r_pr_n[Y_{1}(\hat{\vec{r}}_p)Y_1(\hat{\vec{r}}_n)]^{({2\mu})}. 
\label{eq:Q}
\end{eqnarray}
Here, 
$M_C$ is the mass of the core nucleus, that is, 
$A_Cm$ for $^{18}$F and $A_Cm+m_{\Lambda}$ for 
$^{19}_{\ \Lambda}$F, where $m_{\Lambda}$ is the mass of $\Lambda$ particle. 
In Eq. (\ref{eq:Q}), 
the
effective
 charges of proton and neutron are given as 
\begin{eqnarray}
e^{(p)}=e+e^{(p)}_{\rm pol},\ e^{(n)}=e^{(n)}_{\rm pol}, 
\end{eqnarray}
respectively. Here we have introduced the polarization charge $e_{\rm pol}^{(\tau)}$ 
to protons and neutrons 
to take into account the core polarization effect (in principle 
one may also consider the polarization charge for the $\Lambda$ particle, 
but for simplicity we neglect it in this paper). 
Their values are determined 
so as to reproduce the measured $B(E2)$ values of $1/2^+\to5/2^+$ transitions in 
$^{17}$F ($64.9\pm1.3\ e^2{\rm fm}^4$) and $^{17}$O ($6.21\pm0.08\ e^2{\rm fm}^4$) \cite{Aj83}. 
In our model we calculate them as single-particle transitions 
in $^{17}$F ($^{16}$O+$p$) and in $^{17}$O ($^{16}$O+$n$). 
The resultant 
values are $e^{(p)}_{\rm pol}=0.098e$ and $e^{(n)}_{\rm pol}=0.40e$, which are close 
to the values given in Ref. \cite{SMA04}. 

The $B(E2)$ value from the 3$^+$ state to the 1$^+$ ground state is then 
computed as,
\begin{equation}
B(E2, 3^+\to 1^+)= 
\frac{1}{7}\,
\left|\langle \Psi_{J=1}\|Q_2\|\Psi_{J=3}\rangle\right|^2,
\end{equation}
where $\langle \Psi_{J=1}\|Q_2\|\Psi_{J=3}\rangle$ is the 
reduced matrix element. 
We will compare this with the corresponding value for the 
$^{19}_{~\Lambda}$F nucleus, that is, 
\begin{eqnarray}
&&\frac{1}{7}\,
\left|\langle \Psi_{I=1}\|Q_2\|\Psi_{I=3}\rangle\right|^2 \nonumber \\
&=&\frac{1}{8}\,\left|\langle [\Phi_{I_c}\Psi_{I=1}]^{J=3/2}\|Q_2\|
[\Phi_{I_c}\Psi_{I=3}]^{J=7/2}\rangle\right|^2, 
\label{eq:BE2}
\\
&=&
\frac{1}{6}\,
\left|\langle [\Phi_{I_c}\Psi_{I=1}]^{J=3/2}\|Q_2\|
[\Phi_{I_c}\Psi_{I=3}]^{J=5/2}\rangle\right|^2 \nonumber \\
&&+\frac{1}{6}\,\left|\langle [\Phi_{I_c}\Psi_{I=1}]^{J=1/2}\|Q_2\|
[\Phi_{I_c}\Psi_{I=3}]^{J=5/2}\rangle\right|^2,
\label{eq:BE2-2}
\end{eqnarray}
which is valid in the weak coupling limit 
\cite{Ta01,HiKMM99,IsKDO11sc,IsKDO12} 
(see Appendix B for the derivation). 

\end{subsection}

\label{sec:model}
\end{section}

\begin{section}{RESULTS AND DISCUSSION}

We now numerically solve the three-body Hamiltonians. 
Because $^{18}$F is a well bound nucleus so that the cut-off does not have to be 
high, we use $E_{\rm cut}=10$ MeV. 
We have confirmed that the result does not drastically change even if 
we use a larger value of the cut-off energy, $E_{\rm cut}$= 50 MeV. 
Especially the ratio 
($\approx 0.96$) 
of the 
$B(E2)$ value for $^{19}_{\ \Lambda}$F to that for $^{18}$F 
is quite stable against the cut-off energy. 
We fit the parameters in the proton-neutron pairing interaction, $x_t$ and 
$\alpha_t$, to the energy of the ground state ($-9.75$ MeV) and 
that of the first excited state ($-8.81$ MeV) of $^{18}$F. 
Their values are $x_t=-1.239$ and $\alpha_t=0.6628$ for 
$E_{\rm cut}=10$ MeV.
We use the box size of $R_{\rm box}=30$ fm. 

The obtained level schemes of $^{18}$F and $^{19}_{\ \Lambda}$F 
as well as the $B(E2)$ values are shown in Fig. \ref{fig:lv-s}. 
The $B(E2,3^+\to1^+)$ value is reduced, which indicates that the nucleus 
shrinks by the attraction of $\Lambda$. In fact, as shown in Table \ref{tb:rms}, 
the root mean square (rms) distance between the core and the 
center-of-mass of the two valence nucleons, 
$\langle r^2_{C-pn} \rangle^{1/2}$, and that between the proton and the neutron 
,$\langle r^2_{p-n} \rangle^{1/2}$, slightly decrease by adding $\Lambda$. 

To make the shrinkage effect more visible, we next show the two-particle density. 
The two-particle density $\rho_2({\bf r}_p,{\bf r}_n)$ is defined by 
\begin{eqnarray}
&&\rho_2(\vec{r}_p,\vec{r}_n) \nonumber\\
&&=\sum_{\sigma_p\sigma_n}
\langle\vec{r}_p\sigma_p,\vec{r}_n\sigma_n|\Psi_{IM}\rangle
\langle\Psi_{IM}|\vec{r}_p\sigma_p,\vec{r}_n\sigma_n\rangle,
\label{eq:rho2}
\end{eqnarray}
where $\sigma_p$ and $\sigma_n$ is the spin coordinates of proton and neutron, 
respectively. 
Setting $\hat{\vec{r}}_p=0$, the density is 
normalized as 
\begin{eqnarray}
&&
\int_0^{\infty}4\pi r_p^2dr_p\int_0^{\infty}r_n^2dr_n \nonumber\\
&\times&\int_0^{\pi}2\pi\sin\theta_{pn}d\theta_{pn}\ \rho_2(r_p,r_n,\theta_{pn})=1,
\end{eqnarray}
where 
$\theta_{pn}=\theta_n$ is the angle between proton and neutron. 
In Fig. \ref{fig:density1} we show the ground state density distributions of 
$^{18}$F (the upper panel) and $^{19}_{\ \Lambda}$F (the lower panel) as a function of 
$r=r_p=r_n$ and $\theta_{pn}$. 
They are weighted by a factor of $8\pi^2r^4\sin{\theta_{pn}}$. 
The distribution slightly moves inward after adding a $\Lambda$ particle. 
To see the change 
clearer, we show in Fig. \ref{fig:density-diff} the difference of the 
two-particle densities,   
$\rho_2(^{19}_{\ \Lambda}{\rm F})-\rho_2(^{18}{\rm F})$, for both the 
$1^+$ and the $3^+$ states. 
One can now clearly see that the density distribution is pulled 
toward the origin by additional 
attraction caused by the $\Lambda$ particle both for the $1^+$ and 
the $3^+$ states. 

\begin{figure}
\begin{center}
\setlength\unitlength{1cm}
\begin{picture}(8,3)(0.5,0)
\put(.5,.9){$1^+$}
\put(.5,3.3){$3^+$}
\put(7.1,-.1){$1^+\otimes\Lambda_{s_{1/2}}$}
\put(7.1,1.6){$3^+\otimes\Lambda_{s_{1/2}}$}
\put(1.6,3.5){$0.94$ MeV}
\put(5.2,1.8){$0.66$ MeV}
\put(1.9,.5){$^{18}$F}
\put(5.4,-.5){$^{19}_{\ \Lambda}$F}
\put(2,3.3){\vector(0,-1){2.2}}
\put(5.5,1.6){\vector(0,-1){1.55}}
\put(2.1,2){$15.8\ e^2{\rm fm}^4$}
\put(5.6,.8){$15.2\ e^2{\rm fm}^4$}
\thicklines
\multiput(0,0)(0,2.4){2}{
\put(1,1){\line(1,0){2.5}}
}
\multiput(0,-0.9945)(0,1.685){2}{
\put(4.5,1){\line(1,0){2.5}}
}
\put(3.9,2.4){$\approx$}
\put(3.85,0.45){$\approx$}
\thinlines
 \qbezier( 3.50000, 1.00000)( 3.52083, 0.97928)( 3.54167, 0.95856)
\qbezier( 4.50000, 0.00550)( 4.47917, 0.02622)( 4.45833, 0.04694)
\multiput( 0.00000, 0.00000)( 0.16667,-0.16575){ 6}{
\qbezier( 3.50000, 1.00000)( 3.54167, 0.95856)( 3.58333, 0.91713)
 }
\qbezier( 3.50000, 3.40000)( 3.51786, 3.36947)( 3.53571, 3.33895)
\qbezier( 4.50000, 1.69050)( 4.48214, 1.72103)( 4.46429, 1.75155)
\multiput( 0.00000, 0.00000)( 0.14286,-0.24421){ 7}{
\qbezier( 3.50000, 3.40000)( 3.53571, 3.33895)( 3.57143, 3.27789)
 }
\end{picture}
\end{center}
\caption{The level scheme and the $B(E2)$ values 
for the $^{18}$F and 
$^{19}_{\ \Lambda}$F nuclei. 
For the $^{19}_{\ \Lambda}$F nucleus, 
a sum of the $B(E2)$ values 
for the [3$^+\otimes \Lambda_{s_{1/2}}]^{J=5/2} \to 
[1^+\otimes \Lambda_{s_{1/2}}]^{J=3/2}$ and the 
[3$^+\otimes \Lambda_{s_{1/2}}]^{J=5/2} \to 
[1^+\otimes \Lambda_{s_{1/2}}]^{J=1/2}$ transitions 
is shown, 
which 
corresponds to the $B(E2)$ value from the 3$^+$ to the 1$^+$ states 
in $^{18}$F (see the text for details). 
The excitation energies are shown on the top of 
each state. 
The measured $B(E2)$ value for $^{18}$F is 
$16\pm0.6\ e^2{\rm fm}^4$ \cite{TWC93}.}
\label{fig:lv-s}
\end{figure}
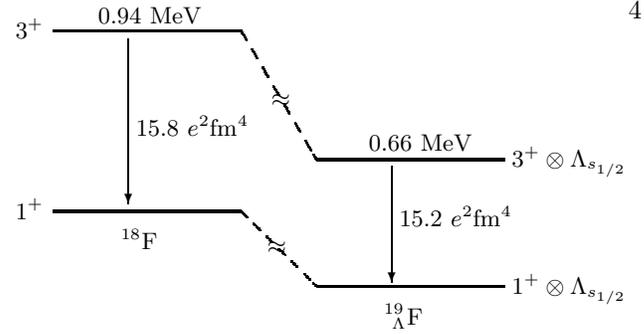

\begin{table}
\caption{The core-pn and p-n rms distances, 
the opening angle between the valence nucleons, and 
the probability for the spin-triplet component 
for the ground and the first excited states 
of $^{18}$F and $^{19}_{\ \Lambda}$F.}
\begin{center}
\begin{tabular}{cc|cccc}
\hline\hline
&& $\langle r^2_{C-pn}\rangle^{1/2}$ & $\langle r^2_{p-n}\rangle^{1/2}$ & $\theta_{pn}$ & $P(S=1)$\\
&& (fm) & (fm) & (deg) & (\%)\\
\hline
$1^+$ & $^{18}$F             & 2.72 & 5.38 & 89.6 & 58.6 \\
$1^+\otimes\Lambda_{s_{1/2}}$  
& $^{19}_{\ \Lambda}$F & 2.70 & 5.33 & 89.5 & 58.4 \\
\hline
$3^+$ & $^{18}$F             & 2.71 & 5.42 & 84.9 & 85.0 \\
$3^+\otimes\Lambda_{s_{1/2}}$  & $^{19}_{\ \Lambda}$F & 2.68 & 5.35 & 84.7 & 85.9 \\
\hline\hline
\end{tabular}
\end{center}
\label{tb:rms}
\end{table}

\begin{figure}
\begin{center}
\includegraphics[scale=.33 ,angle=-90]{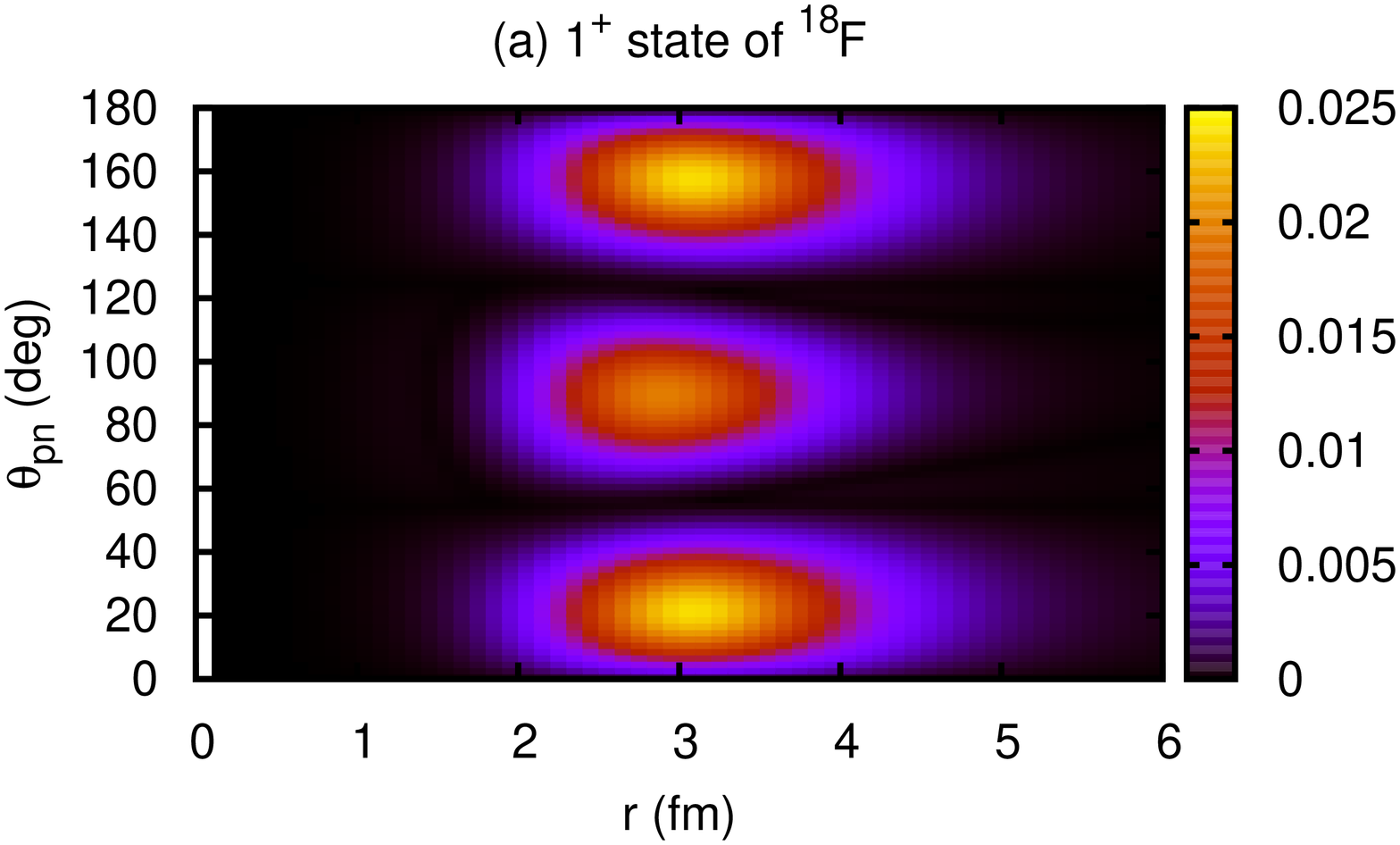}
\includegraphics[scale=.33 ,angle=-90]{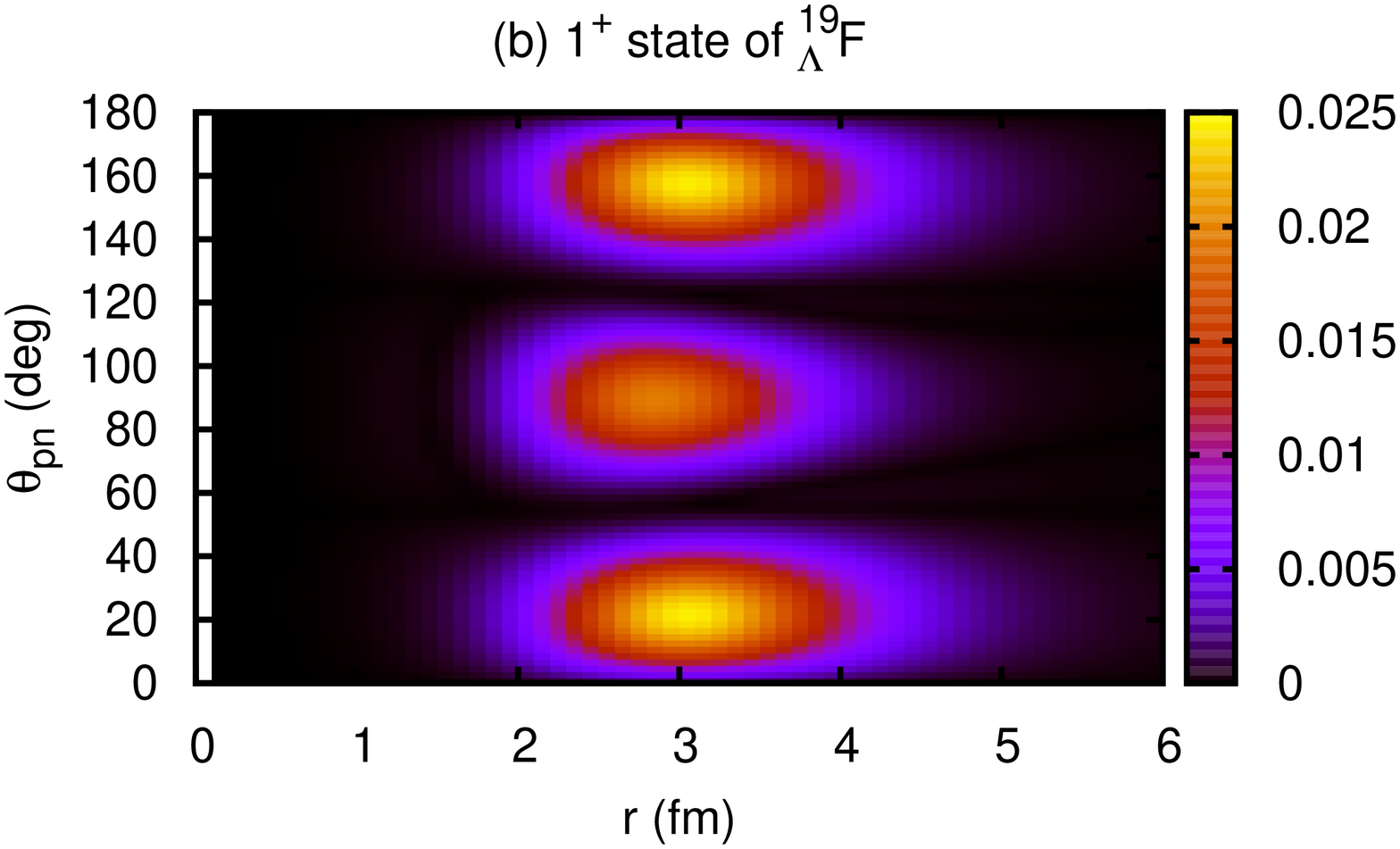}
\end{center}
\caption{(Color online) The two-particle densities of the ground state of   
(a) $^{18}$F and (b) $^{19}_{\ \Lambda}$F as a function of 
$r_p=r_n\equiv r$ and the opening angle between the proton and the 
neutron, $\theta_{\rm pn}$. Those densities are multiplied by a weight 
factor of $8\pi^2r^4\sin\theta_{\rm pn}$. }
\label{fig:density1}
\end{figure}

\begin{figure}
\begin{center}
\includegraphics[scale=.33 ,angle=-90]{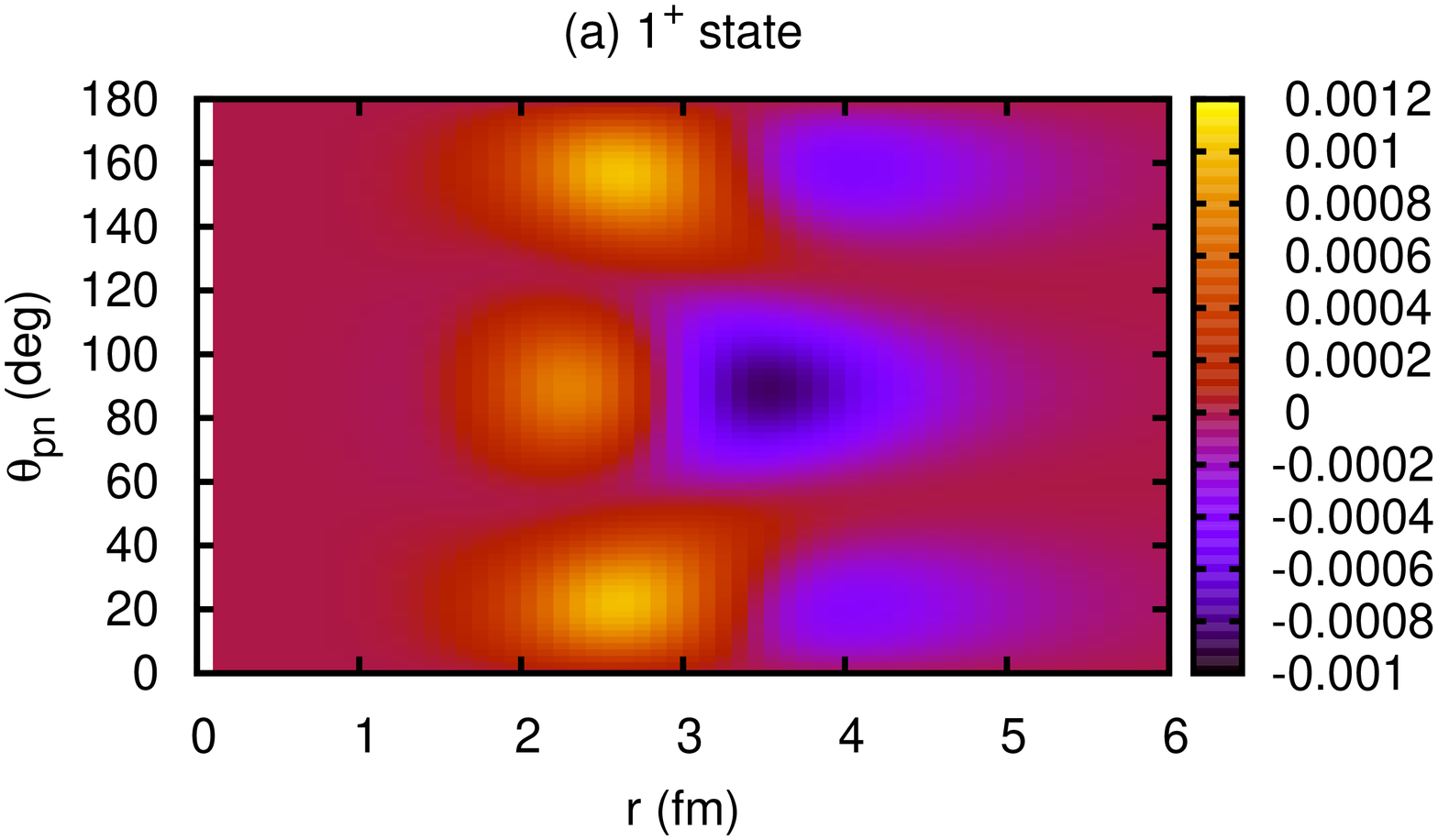}
\includegraphics[scale=.33 ,angle=-90]{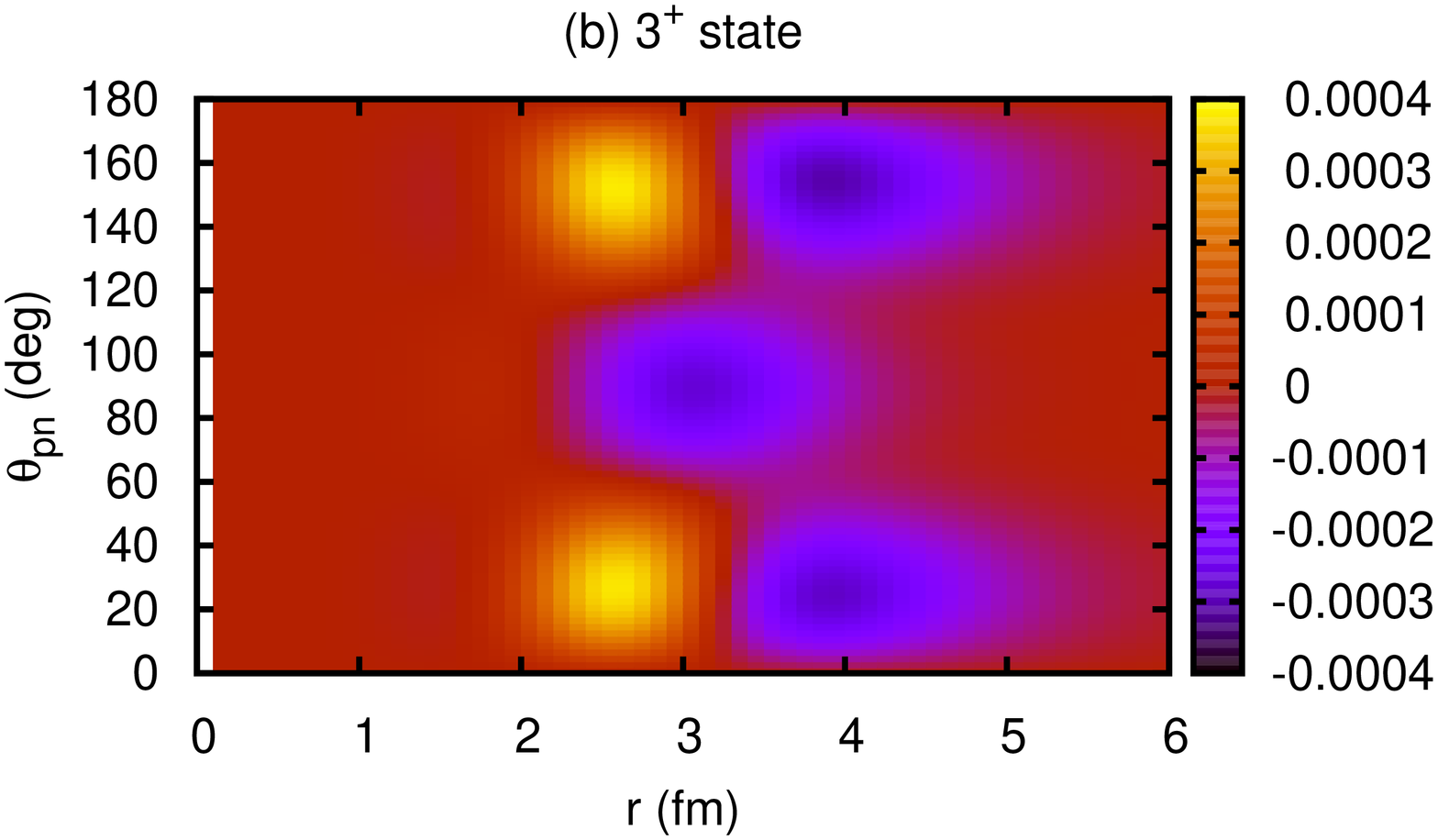}
\end{center}
\caption{(Color online) The difference of the 
density distribution between $^{19}_{\ \Lambda}$F and $^{18}$F for 
(a) the $1^+$ state and (b) the $3^+$ state. }
\label{fig:density-diff}
\end{figure}

Let us next discuss the change in excitation energy. 
As shown in Fig. 1, it is deceased by addition of $\Lambda$,  
similar to $^{6}$Li and $^{7}_{\Lambda}$Li. 
In order to clarify the mechanism of this 
reduction we estimate the energy
  gain of each valence configuration, 
treating $V_{\Lambda}(r_p)+V_{\Lambda}(r_n)=\Delta V$ by 
the first order perturbation theory: 
\begin{eqnarray}
\Delta E_I&=&\langle\Psi^{(IM_I)}_{^{18}{\rm F}}|\Delta V|\Psi^{(IM_I)}_{^{18}{\rm F}}\rangle \nonumber\\
&=&\sum_{j_{\alpha}\ell_{\alpha}}\sum_{j_{\beta}\ell_{\beta}}
\sum_{n_{\alpha}n_{\alpha'}}\sum_{n_{\beta}n_{\beta'}}
C_{n_{\alpha'}\ell_{\alpha}j_{\alpha},n_{\beta'}\ell_{\beta}j_{\beta}}^{I*}
C_{n_{\alpha}\ell_{\alpha}j_{\alpha},n_{\beta}\ell_{\beta}j_{\beta}}^{I} \nonumber\\
&\times&\biggl[\biggr.
\delta_{n_{\beta'}n_{\beta}}\int_0^{\infty}r_p^2dr_p\ 
\phi^{(p)}_{n_{\alpha'}\ell_{\alpha}j_{\alpha}}(r_p)^*\,V_{\Lambda}(r_p)
\phi^{(p)}_{n_{\alpha}\ell_{\alpha}j_{\alpha}}(r_p) \nonumber\\
&\biggl. +&\delta_{n_{\alpha'}n_{\alpha}}\int_0^{\infty}r_n^2dr_n\ 
\phi^{(n)}_{n_{\beta'}\ell_{\beta}j_{\beta}}(r_n)^*\,V_{\Lambda}(r_n)
\phi^{(n)}_{n_{\beta}\ell_{\beta}j_{\beta}}(r_n)\biggr] \nonumber\\
&\equiv&\sum_{j_{\alpha}\ell_{\alpha}j_{\beta}\ell_{\beta}}
\Delta\epsilon^{(I)}_{j_{\alpha}\ell_{\alpha}j_{\beta}\ell_{\beta}}, 
\end{eqnarray}
where $\Delta\epsilon^{(I)}_{j_{\alpha}\ell_{\alpha}j_{\beta}\ell_{\beta}}$ is the contribution of each 
configuration to the total approximate energy change $\Delta E_I$. 
We show in the first and the second columns in Table \ref{tb:comp} 
dominant 
valence configurations and their occupation probabilities. 
In the third and the 
fourth columns of the Table \ref{tb:comp} 
are the energy gains of each configuration 
$\Delta\epsilon$ and $\Delta\epsilon/P$, respectively, 
where $P$ is the occupation probability in the $^{18}$F nucleus of the 
corresponding configuration. 
Notice that $\Delta\epsilon/P$ is the largest for 
the $s\otimes s$ configuration, and the second largest for the 
$s\otimes d$ and $d\otimes s$, 
because $s$ wave states have more overlap with the $\Lambda$ occupying the 
$1s_{1/2}$ orbital. 
In the ground state of $^{18}$F both proton and neutron occupy $s$-wave states with a 
probability of 11.63 \%, while in the excited $3^+$ state one of the two valence nucleons 
occupies an $s$-wave state with a probability of 55.71 \%. 
Thus the $3^+$ state has much more $s$-wave component than 
the $1^+$ state. 
Therefore the 
valence nucleons in the $3^+$ state have more 
overlap with the $\Lambda$ particle and gain more binding compared to the 
ground state. In fact, as one can see in Table \ref{tb:comp}, the probabilities of the 
configurations with $s$-wave grow up by adding $\Lambda$. 

We have repeated the same calculations by turning off 
the core-nucleon spin-orbit interaction in Eq.(\ref{eq:WS}), 
which is the origin of the core-deuteron spin-orbit interaction. We have confirmed that the excitation 
energy still decreases without the spin-orbit interaction. 
We have also carried out calculations for the ground ($0^+$) and the
first excited ($2^+$) levels in $^{18}$O and
$^{19}_{\ \Lambda}$O with only the spin-singlet (iso-triplet) channel
interaction.
In this case the spin-orbit interaction between the core and two
neutrons is absent.
In this case also the excitation energy decreases by adding a $\Lambda$ 
particle.
Therefore we find that the mechanism 
of the reduction of the excitation energy in $^{18}$F 
is indeed different from the case of lithium,  
where the $LS$ interaction between the core and deuteron 
plays an important role in the latter.

\begin{table}
\caption{Dominant configurations and their occupation probabilities for 
the $1^+$ and $3^+$ states of $^{18}$F and $^{19}_{~\Lambda}$F. 
The energy gains of each configuration estimated by the first order 
perturbation theory are also shown in the third and fourth columns, where 
$P$ is the occupation probability in the $^{18}$F nucleus. }

\begin{center}
\begin{tabular}{c|cccc}
\hline\hline
$1^+$ state &  & & &  \\
\hline 
Configuration &\multicolumn{2}{|c}{Occupation probability}& $\Delta\epsilon$ & $\Delta\epsilon/P$ \\
 & $^{18}$F & $^{19}_{\ \Lambda}$F & (MeV) & (MeV)\\
\hline
$\pi_{d_{5/2}}\otimes \nu_{d_{5/2}}$ & 53.69 \% & 54.29 \% & $-0.38$ & $-0.71$\\
$\pi_{d_{3/2}}\otimes \nu_{d_{5/2}}$ & 15.85 \% & 15.02 \% & $-0.10$ & $-0.63$\\
$\pi_{d_{5/2}}\otimes \nu_{d_{3/2}}$ & 15.41 \% & 14.57 \% & $-0.10$ & $-0.65$\\
$\pi_{s_{1/2}}\otimes \nu_{s_{1/2}}$ & 11.63 \% & 12.76 \% & $-0.13$ & $-1.12$\\
\hline\hline
$3^+$ state &  & & &  \\
\hline
$\pi_{d_{5/2}}\otimes \nu_{d_{5/2}}$ & 38.30 \% & 35.98 \% & $-0.27$ & $-0.70$\\
$\pi_{s_{1/2}}\otimes \nu_{d_{5/2}}$ & 28.30 \% & 29.64 \% & $-0.26$ & $-0.92$\\
$\pi_{d_{5/2}}\otimes \nu_{s_{1/2}}$ & 27.41 \% & 28.68 \% & $-0.26$ & $-0.95$\\
\hline\hline
\end{tabular}
\end{center}
\label{tb:comp}
\end{table}

\label{sec:result}
\end{section}

\begin{section}{SUMMARY}
We have calculated the energies of the lowest two levels and $E2$ 
transitions of $^{18}$F and 
$^{19}_{\ \Lambda}$F using a simple three-body model. 
It is found that $B(E2,3^+\to1^+)$ is slightly 
reduced, as is expected from the 
shrinkage effect of $\Lambda$. We have indeed seen that the distance between the valence 
two nucleons and the $^{16}$O core decreases after adding a $\Lambda$ particle. 
We also found that excitation energy of the 
$3^+$ state is decreased. We observed 
that the $3^+$ state has much more $s$-wave component 
than the ground state 
and thus gains more binding coupled 
with the $\Lambda$ occupying $1s_{1/2}$ orbital. 
This leads to a conclusion that 
the excitation 
energy of the first core excited state $3^+\otimes \Lambda_{s_{1/2}}$ of $^{19}_{\ \Lambda}$F 
becomes smaller than the corresponding excitation in $^{18}$F. We have 
pointed out that the mechanism of this 
reduction is different from that of $^6$Li and $^7_{\Lambda}$Li, 
where the
 core-deutron spin-orbit interaction plays an important role \cite{HK11}.
This may suggest that the information on the wave function of
a core nucleus can be studied using the spectroscopy of $\Lambda$
hypernuclei.

In this paper, we used a spin-independent $N$-$\Lambda$ interaction. 
To be more realistic and quantitative, 
it is an interesting future work to 
employ a spin dependent $N$-$\Lambda$ 
interaction and explicitly take into account the core excitation. 
It may also be important to explicitly take into account 
the tensor correlation between 
the valence proton and neutron. 

\label{sec:summary}
\end{section}

\begin{acknowledgments}
We thank T. Koike, T. Oishi, 
and H. Tamura for useful discussions. 
The work of Y. T. was supported by the Japan Society for the Promotion of 
Science for Young Scientists. 
This work was also supported by Grant-in-Aid for JSPS Fellows under 
the program number 24$\cdot$3429 and the Japanese
Ministry of Education, Culture, Sports, Science and Technology
by Grant-in-Aid for Scientific Research under
the program number (C) 22540262. 
\end{acknowledgments}

\appendix
\section{Matrix elements of $V_{pn}$}

In this appendix we explicitly give an expression for the 
matrix elements of the 
proton-neutron pairing interaction $V_{pn}$ given by Eq. (\ref{eq:pairing}). 
They read 
\begin{eqnarray}
&&\langle\alpha'\beta',IM|V_{pn}|\alpha\beta,IM\rangle \nonumber\\
&=&
\langle\alpha'\beta',IM|
\bigl[\hat{P}_s\delta^{(3)}(\vec{r}_p-\vec{r}_n)g_s(r_p)+\bigr. \nonumber\\
&&\hspace{1.5cm}\bigl.\hat{P}_t\delta^{(3)}(\vec{r}_p-\vec{r}_n)g_t(r_p)\bigr]
|\alpha\beta,IM\rangle,
\end{eqnarray}
where we have defined 
\begin{eqnarray}
g_{s}(r)=v_{s}\biggl[1+x_s\biggl(\frac{1}{1+e^{(r-R)/a}}\biggr)^{\alpha_s}\biggr],
\end{eqnarray}
and similarly for $g_t(r)$. By rewriting the basis into the $LS$-coupling 
scheme one obtains 
\begin{widetext}
\begin{eqnarray}
&&({\rm the~singlet\ term}) \nonumber\\
&=&
\frac{(-)^{\ell_{\alpha}+j_{\beta}-\ell_{\alpha'}-j_{\beta'}}}{8\pi}
\hat{j}_{\alpha}\hat{j}_{\alpha'}\hat{j}_{\beta}\hat{j}_{\beta'}
\hat{\ell}_{\alpha}\hat{\ell}_{\alpha'}\hat{\ell}_{\beta}\hat{\ell}_{\beta'}
\left\{\begin{array}{ccc}
j_{\alpha} & j_{\beta} & I\\ 
\ell_{\beta} & \ell_{\alpha} & \frac{1}{2}
\end{array}\right\}
\left\{\begin{array}{ccc}
j_{\alpha'} & j_{\beta'} & I\\ 
\ell_{\beta'} & \ell_{\alpha'} & \frac{1}{2}
\end{array}\right\}
\left(\begin{array}{ccc}
\ell_{\alpha} & \ell_{\beta} & I\\ 
0 & 0 & 0
\end{array}\right)
\left(\begin{array}{ccc}
\ell_{\alpha'} & \ell_{\beta'} & I\\ 
0 & 0 & 0
\end{array}\right)
 \nonumber\\
&&\times
\int_0^{\infty}r^2dr\ \phi^{(p)}_{\alpha'}(r)^*\phi^{(n)}_{\beta'}(r)^*
\phi^{(p)}_{\alpha}(r)\phi^{(n)}_{\beta}(r)g_s(r), 
\end{eqnarray}
and
\begin{eqnarray}
&&({\rm the~triplet\ term}) \nonumber\\
&=&
\frac{3}{4\pi}
\hat{j}_{\alpha}\hat{j}_{\alpha'}\hat{j}_{\beta}\hat{j}_{\beta'}
\hat{\ell}_{\alpha}\hat{\ell}_{\alpha'}\hat{\ell}_{\beta}\hat{\ell}_{\beta'}
\sum_{L}\hat{L}^2
\left\{\begin{array}{ccc}
\ell_{\alpha} & \ell_{\beta} & L\\
\frac{1}{2} & \frac{1}{2} & 1\\
j_{\alpha} & j_{\beta} & I
\end{array}\right\}
\left\{\begin{array}{ccc}
\ell_{\alpha'} & \ell_{\beta'} & L\\
\frac{1}{2} & \frac{1}{2} & 1\\
j_{\alpha'} & j_{\beta'} & I
\end{array}\right\}
\left(\begin{array}{ccc}
\ell_{\alpha} & \ell_{\beta} & L\\ 
0 & 0 & 0
\end{array}\right)
\left(\begin{array}{ccc}
\ell_{\alpha'} & \ell_{\beta'} & L\\ 
0 & 0 & 0
\end{array}\right) \nonumber\\
&&\times
\int_0^{\infty}r^2dr\ \phi^{(p)}_{\alpha'}(r)^*\phi^{(n)}_{\beta'}(r)^*
\phi^{(p)}_{\alpha}(r)\phi^{(n)}_{\beta}(r)g_t(r), 
\end{eqnarray}
where $\hat j=\sqrt{2j+1}$. From these equations, 
it is apparent that for odd (even) $I$ and even (odd) 
parity states, such as 1$^+$ and 3$^+$, 
the singlet term always vanishes because 
\begin{equation}
\left(\begin{array}{ccc}
\ell_{\alpha} & \ell_{\beta} & I\\ 
0 & 0 & 0
\end{array}\right)=0,
\end{equation}
for $\ell_\alpha+\ell_\beta+I$ = odd. 

\end{widetext}

\section{Extraction of the core transition from $B(E2)$ values 
of a hypernucleus}

We consider a hypernucleus with a $\Lambda$ particle weakly coupled 
to a core nucleus. 
In the weak coupling approximation, the wave function for the 
hypernucleus with angular momentum $J$ and its $z$-component $M$ 
is given by 
\begin{eqnarray}
|JM\rangle &=& [|I\rangle\otimes|j_\Lambda\rangle]^{(JM)} \nonumber \\
&=&\sum_{M_I,m_\Lambda}
\langle I M_I j_\Lambda m_\Lambda|JM\rangle |IM_I\rangle |j_\Lambda m_\Lambda\rangle,
\end{eqnarray}
where $|IM_I\rangle$ and $|j_\Lambda m_\Lambda\rangle$ are the wave functions 
for the core nucleus and the $\Lambda$ particle, respectively. 
Suppose that the operator $\hat{T}_{\lambda\mu}$ 
for electromagnetic transitions is independent of the $\Lambda$ particle. 
Then, the square of the reduced matrix element of $\hat{T}_{\lambda\mu}$ between 
two hypernuclear states reads (see Eq. (7.1.7) in Ref. \cite{Edmonds} 
as well as Eq. (6-86) in Ref. \cite{BM75}),
\begin{eqnarray}
|\langle J_f\|T_\lambda\|J_i\rangle |^2
&=& (2J_i+1)(2J_f+1)
\left\{\begin{array}{ccc}
I_f & J_f & j_\Lambda \\
J_i & I_i & \lambda 
\end{array}\right\}^2 \nonumber \\
&&\times\langle I_f\|T_\lambda\|I_i\rangle^2.
\end{eqnarray}
Notice the relation (see Eq. (6.2.9) in Ref.\cite{Edmonds}),
\begin{equation}
\sum_{J_f} (2J_f+1)
\left\{\begin{array}{ccc}
I_f & J_f & j_\Lambda \\
J_i & I_i & \lambda 
\end{array}\right\}^2 =\frac{1}{2I_i+1}.
\end{equation}
This yields 
\begin{eqnarray}
\sum_{J_f}B(E\lambda; J_i\to J_f) &=&
\sum_{J_f}\frac{1}{2J_i+1}
|\langle J_f\|T_\lambda\|J_i\rangle |^2 \nonumber\\
&=&
\frac{1}{2I_i+1}\langle I_f\|T_\lambda\|I_i\rangle^2,
\end{eqnarray}
which is nothing but the $B(E\lambda)$ value of the core nucleus 
from the state $I_i$ to the state $I_f$. 
This proves Eqs. (\ref{eq:BE2}) and (\ref{eq:BE2-2}) in Sec. II 
for the specific case of $I_i=3$ and $I_f=1$.

\end{document}